# Measuring Technological Distance for Patent Mapping

**Bowen Yan**

SUTD-MIT International Design Centre
& Engineering Product Development Pillar
Singapore University of Technology and Design
8 Somapah Road, Singapore 487372
Email: bowen_yan@sutd.edu.sg

**Jianxi Luo**[*]

SUTD-MIT International Design Centre
& Engineering Product Development Pillar
Singapore University of Technology and Design
8 Somapah Road, Singapore 487372
Email: luo@sutd.edu.sg

[*] Corresponding author

**Abstract**

Recent works in the information science literature have presented cases of using patent databases and patent classification information to construct network maps of technology fields, which aim to aid in competitive intelligence analysis and innovation decision making. Constructing such a patent network requires a proper measure of the distance between different classes of patents in the patent classification systems. Despite the existence of various distance measures in the literature, it is unclear how to consistently assess and compare them, and which ones to select for constructing patent technology network maps. This ambiguity has limited the development and applications of such technology maps. Herein, we propose to compare alternative distance measures and identify the superior ones by analyzing the differences and similarities in the structural properties of resulting patent network maps. Using United States patent data from 1976 to 2006 and International Patent Classification system, we compare 12 representative distance measures, which quantify inter-field knowledge base proximity, field-crossing diversification likelihood or frequency of innovation agents, and co-occurrences of patent classes in the same patents. Our comparative analyses suggest the patent technology network maps based on *normalized co-reference* and *inventor diversification likelihood* measures are the best representatives.

# 1. Introduction

To pursue innovation, inventors, companies or R&D organizations, cities or countries continually diversify to explore technology fields different from their past ones, or combine their existing knowledge with those of new fields to build new technological capabilities (Schumpeter, 1934; Dosi, 1982). Therefore, innovation can be viewed as a process of searching and combining knowledge across different technology fields. The variety of technology fields together constitutes the "technology space", in which the fields may have different distances between each other (Teece et al., 1994; Breschi et al., 2003; Kay et al., 2014). In turn, the heterogeneous structure of the technology space may condition the diversification paths or knowledge recombination prospects of innovation agents (e.g., Thomas Edison, Google, China) with different knowledge positions in the space, and condition the development potentials of specific technologies (e.g., fuel cells, robots, aircrafts) given the positions of their knowledge base in the whole technology space.

Recent studies have proposed to represent the technology space as a network map of technology fields based on mining patent data (Leydesdorff et al., 2014; Kay et al., 2014; Nakamura et al., 2014). In such a network, a vertex represents a technology field and is operationalized as a patent technology class. The weighted edge between a pair of vertices denotes the distance between the vertex-represented technology fields. One can also overlay such a network map by highlighting a subset of fields that are associated with a technological design domain of interest (e.g., robotics, fuel cells), or the innovative activities of an innovation agent (Kay et al., 2014). Fig. 1 illustrates an example of the network overlaid with a subset of highlighted fields where Google Inc. had been granted US patents over time. Such an overlay map locates the fields where a specific agent has developed innovation capabilities and also reveals the evolution of such capabilities.

Assessment of the relative network positions of the subset of fields on the overlaid map may illuminate new fields that are proximate to them in the technology space and present great knowledge recombination potential with them (Fleming, 2001; Nakamura et al., 2014). Such analyses may also lead to insights on the directions and paths of technology diversification of an innovation agent, e.g., firm, city or country (Breschi et al., 2003; Rigby, 2013; Boschma et al., 2014), or help forecast development directions and potentials of an emerging technology, e.g., fuel cells or solar cells (Kajikawa et al., 2008; Ogawa and Kajikawa, 2015; Benson and Magee, 2013; 2015), given the locations of its established knowledge base in the technology space. In general, such a map of technology fields will be useful to aid in technology road mapping of innovation agents or technology-based industries and forecasting development directions of emerging technologies.

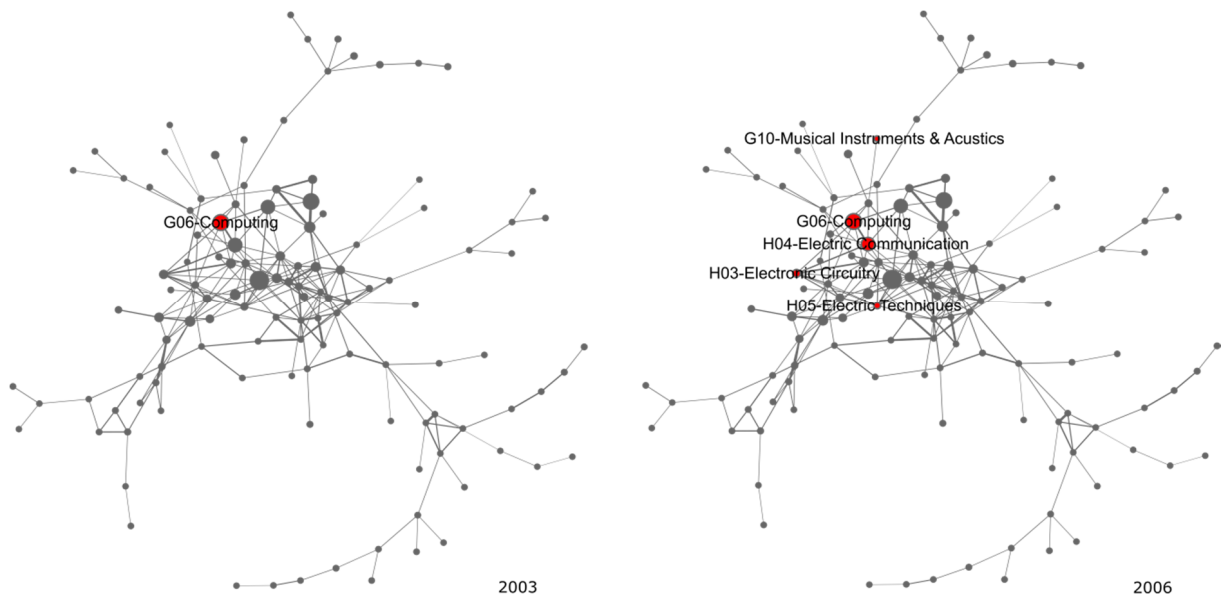

**Fig. 1.** Example of the technology network map overlaid with Google's knowledge positions.[1] Each vertex represents a 3-digit technology class defined in the International Patent Classification (IPC) system, and its size corresponds to the total number of patents in the class from 1976 to 2006. An edge between pairs of vertices is measured as inventor cross-field diversification likelihood (see details of the measure in Section 2 and Section 3.2). The original network is extremely dense. The network visualized here contains the maximum spanning tree (i.e., a minimal set of edges that connect all vertices and maximize total edge weights) as the backbone plus the strongest edges, which together make the total number of edges be twice of the vertices, as suggested by Hidalgo et al. (2007) for best visualization. A vertex is highlighted in red color if Google had patents in the corresponding technology class in a given time period. Details about this network are in Section 4.1.

For such a network to adequately represent the technology space requires an appropriate measure of the distance between different technology fields (Jaffe, 1986; Joo and Kim, 2010; Altuntas et al., 2015). Although various distance measures have been proposed from different perspectives in the literature (see a review of the measures in Section 2), they have not been assessed and compared using consistent criteria or methodology. It is unclear which measures are superior for the purpose of constructing technology network maps, and which measures are more representative of the others. This ambiguity has limited the use of technology network maps in technology forecasting and roadmap analyses.

In this paper, we recognize that the choice of inter-field distance measures determines the structure of the technology network to be constructed, which in turn influences the innovation-related insights to be drawn from the network analysis. Following this logic, we propose a strategy to use the structures of the overall technology networks resulting from the choices of distance measures as the lens or protocol to assess and compare corresponding measures. We also demonstrate this strategy through a comparative analysis of twelve alternative distance measures, by using a few network metrics and patent data from United States Patent and Trademark Office (USPTO). These twelve measures are chosen as representatives of the main types of distance measures in the literature. Our analyses yield new understandings on the differences and similarities of these measures, and also shed light on some of these measures that yield more representative network maps than others for constructing technology network maps.

---

[1] Fig. 1 highlights that Google's first patents were in IPC class G06 *computing* in 2003. Later, Google diversified into additional fields. In 2006, it had patents in H03 *electronic circuitry*, H04 *electronic communication* and H05 *electric techniques*, and G10 *music instruments and acoustics*, in addition to G06. These new fields appear to be proximate to the original field G06, within the core of the technology network map.

The paper is organized as follows. We first survey the literature on various quantitative measures of the distance of technology fields in Section 2. Section 3 introduces our data, methodology and twelve distance measures. Section 4 reports and discusses results. Section 5 concludes the paper with suggestions for future work.

## 2. Literature review: Measures of distance between technology fields

In the literature, various technological distance measures have been developed, using the information of references, classifications and inventor identities in patent documents. Some of these measures were noted as "technological proximity", which is direct reverse concept of technological distance (Jaffe, 1986; Leydesdorff et al., 2014).

*2.1 Patent reference-based measures*

One strand of the measures uses patent citation information to calculate indicators of knowledge distance of different technology fields. For instance, to construct the network map of IPC classes, Leydesdorff et al. (2014) used the *cosine similarity index* to normalize the citing-to-cited relationships between technology classes in an aggregated citation matrix. The angular cosine value of the two vectors of citations from two classes to other classes captures the similarity of their knowledge bases. Kay et al. (2014) also used the cosine similarity as the measure of technological distance among different patent categories, some of which combine original IPC patent classes to optimize the size distribution of classes for the sake of visualization. Indeed, Jaffe (1986) was the first to propose this index for technology mapping, whereas he used it to measure the correlation between the vectors representing the distributions of firms' patents in a set of technology fields.

In addition, to measure the knowledge distance between patents, *co-citations*, i.e., the number of shared forward citations of two patents, and *bibliographic coupling*, i.e., the number of shared backward citations (i.e. references) of two patents, were popularly used (Iwan von Wartburg et al., 2005; Leydesdorff and Vaughan, 2006). A co-citation index can be further normalized over the total number of citations for each article, i.e., the *Jaccard index* (Small, 1973), or over a probabilistic measure of expected co-citation counts (Zitt et al., 2000). The formulas of co-citations or bibliographic coupling of patents or academic articles can be adopted to measure and indicate the distance of different technology classes.

*2.2 Patent classification-based measures*

Scholars have also used the "co-classification" information of patents to develop indicators of the distance between technology classes. A patent belongs to at least one, but usually multiple classes assigned by the patent examiners of the issuing offices. Using this information, the distance between technology fields can be indicated by the co-occurrence of classification codes assigned to individual patent documents (Engelsman and van Raan, 1994). The assumption is that the frequency in which two classes are jointly assigned to the same patents will be high if these two classes are proximate. This assumption is similar to the *survivor principle* in economics (Stigler, 1968), which suggests that surviving firms' behaviors are more observable in empirical data, because they are more efficient and thus make firms survive and observable.

Jaffe (1986) was also the first to apply the *cosine index* to measuring the distance of firms' technological portfolios based on the symmetrical matrix of the frequency of two technology classes being jointly assigned to the same patent that belongs to the observed firms. Later, the cosine index was adopted for the general symmetrical *co-occurrence matrix* in which each cell represents the total number of patents that are assigned with both technology fields represented by the respective row and column (Breschi et al., 2003; Ejermo, 2005; Kogler et al., 2013). Leydesdorff and Vaughan (2006) argued that the symmetrical co-occurrence matrix contains similarity data and can be analyzed directly, whereas further normalization of the co-occurrence matrix using the Pearson correlation or cosine may distort the data and generate spurious correlations.

Leydesdorff (2008) further proposed to analyze the asymmetrical classification assignment matrix, with patents as the units of analysis and the technology classes as the column variables, and use the cosine index to associate the column variables. He also found that networks built using classification data match poorly with those generated by citation data, and the classification data might be less useful than co-citation data for technology network mapping, primarily because the classifications were assigned poorly by the ISI staff. In addition, Joo and Kim (2010) also argued that co-classification measures may not directly assess the distance of technology fields and proposed to create a multi-dimensional contingency table to represent patent classification data and apply the Mantel-Haenszel common odds ratio on the table for measurement.

Furthermore, Nesta and Dibiaggio (2005), using the typical co-occurrence matrix, measured the deviation of the number of observed patents shared between classes from the expected number of randomly shared patents, in order to reveal the distance among fields, following Teece et al. (1994) who initially developed this normalization method to measure the distance between industrial fields. This measure takes a *t*-statistic form and adjusts for the effects of class sizes. Similarly, recognizing the uneven importance of different patents and their classes, Altuntas et al. (2015) used the data of forward citations of each patent and of the size of each technology class in terms of total patent count to weight each patent occurring between a pair of technology classes, when counting the occurrences of the same patents in a pair of classes.

*2.3 Likelihood of diversification as measures of distance*

Another group of measures utilizes the data on field-crossing diversification behaviors of innovation agents (e.g., countries, regions, cities, organizations or inventors) to indicate the proximity (the reverse of distance) between technology fields. In studying the product space, Hidalgo et al. (2007) measured the proximity between two product categories in terms of the likelihood for an average country to develop strong relative comparative advantage (RCA) in one product category, given that it has developed strong RCA in the other. The assumption is that this likelihood is high if the capabilities required to produce products in one category are similar to those required to produce another product. In other words, the likelihood of the diversification of countries across two product categories may indicate the proximity of the knowledge base of these two product categories. Boschma et al. (2013) applied the same proximity measure to studying the diversification of productive capabilities of different regions in Spain based on export product categories.

Although the studies of product space were based on export and import data and the custom classifications of products, their proximity measure can be adapted to patent data and patent technology classifications. For instance, a mathematically similar index called the revealed technological advantage (RTA) has been used to measure the pattern of technological specialization of innovation agents (Cantwell and Vertova, 2004; Hall et al., 2001). Boschma et al. (2014) applied this measure to calculating the likelihood of technology diversification at the region and city levels, and used such a likelihood as edge weight in the network of patent technology classes.

In parallel, Teece et al. (1994) estimated how much the frequencies that firms diversify in combinations of 4-digit SIC industries deviate from what one would expect if diversification patterns were random. They called it a "survivor-based measure", because their inspiration was from the *survivor principle* in economics (Stigler, 1968), which suggests that efficient firms survive and contribute to empirical observations and regularity. Following the survivor principle, Teece et al. argued that one can observe that firms diversify more often across industries that are more proximate, so that the number of diversifying firms in a pair of industries may indicate the distance of the industries. Particularly, this measure is superior in that it extracts the information about the true distance in the number of empirical observations by comparing it to the expected value under the hypothesis that diversification is random and not affected by the true distance. In doing so, it adjusts for industry size such that it can be compared consistently across industry pairs. Despite being initially developed to measure industry distance, this measure can be easily leveraged to measure the distance of technology fields, based on patent data.

*2.4 Other measures*

In addition to the information on references, classifications, and inventor identities in patent documents, patent texts have also been analyzed to measure the distance of different technologies and fields. For instance, Nakamura et al. (2014) measured the technological distance of patents in the sub-domains of automobile and aircraft industries by using cosine similarity of the vectors representing occurrence frequencies of words in the patent titles and abstracts of pairs of sub-domains. Fu et al. (2013) proposed a technological distance measure as the text similarity between patents in terms of the functional meanings of the verbs that appear in the description texts of patents. Information of "functional similarity" is useful, because those different solutions or mechanisms used in different inventions to address similar functions present great potential to be recombined into new technologies.

Despite the variety of distance measures in the literature, they have not been assessed and compared using a consistent methodology or criterion. To address this gap, this paper presents a strategy and methodology to assess and compare alternative measures by investigating the similarities and differences in the structures of their resulting technology networks. To implement this strategy, we analyze twelve distance measures that belong to the categories of measures reviewed in Section 2.1, 2.2 and 2.3, respectively.

## 3. Data and methodology

*3.1 Data*

The vertices in our technology network maps are patent classes defined in the International Patent Classification (IPC) system, following many other authors who have considered IPC classes the most suitable and stable representations of technology fields (Leydesdorff et al., 2014).[2] The IPC system includes 8 broad technical domains, which can be subdivided into, for example, 3-digit and 4-digit level subclasses. For the best visualization without losing necessary details and resolution of the technology landscape, we chose 3-digit classes to represent vertices in networks. Some undefined classes, for example, "A99 - subject matter not otherwise provided for in this section," are excluded from the analysis. As a result, the networks contain 121 vertices, i.e. 3-digit level IPC classes. We use the patent data from 1976 to 2006 from United States Patent & Trademark Office (USPTO) and NBER Patent Data Project[3]. The data set contains 3,186,310 utility patents. Each patent is classified in one or multiple IPC classes.

*3.2 Distance measures*

The literature review has shed light on at least four categories of distance measures:

1) the proximity (either similarity or relatedness) of knowledge bases of the innovation activities in a pair of technology fields, using patent citation data;
2) the likelihood for the same innovation agents (i.e., inventors, R&D organizations, or countries) to invent technologies in a pair of technology fields, using data on the successful patenting records of the agents (i.e., in which classes one has patents);
3) the frequency to observe the same innovation agents inventing technologies in a pair of technology fields, using data on the successful patenting records of the agents;
4) the frequency to observe a pair of technology fields being assigned to the same patents, using data on the co-classifications of patents.

In this paper, we choose to analyze 3 specific measures that are most representative for each above category, totaling 12 measures. Table 1 summarizes these 12 measures that follow respective rationales.

---

[2] This paper focuses on comparing vertex distance measures, so vertex definitions are fixed. Future research may conduct similar analyses based on different definitions of vertices, such as United States Patent Classes and recently proposed hybrid patent categories (Kay et al. 2014; Bensen and Magee, 2013; 2015).
[3] NBER Patent Data Project website: https://sites.google.com/site/patentdataproject/Home

Table 1 Twelve distance measures in four different categories

| Rationale | Data required | Measures | Definitions |
| --- | --- | --- | --- |
| A: Similarity or relatedness of knowledge bases | Patent references | A1: Normalized co-reference | The count of shared citations, normalized by the count of all unique citations of patents in a pair of classes |
| | | A2: Class-to-class cosine similarity | The cosine of the angle of the two vectors representing two technology classes' distributions of citations into all patent classes |
| | | A3: Class-to-patent cosine similarity | The cosine of the angle of the two vectors representing two technology classes' distributions of citations into unique patents |
| B: Likelihood for innovation agents to diversify across fields | Bibliographical information of inventors, assignees and regions | B1: Inventor diversification likelihood | Minimum of the pairwise conditional probabilities of an inventor having stronger than average patenting records in one class, given that he also has stronger than average records in the other |
| | | B2: Organization diversification likelihood | Minimum of the pairwise conditional probabilities of an organization having stronger than average patenting records in one class, given that it also has stronger than average records in the other. |
| | | B3: Country diversification likelihood | Minimum of the pairwise conditional probabilities of a country having stronger than average patenting records in one class, given that it also has stronger than average records in the other |
| C: Frequency to observe innovation agents diversifying across fields | Bibliographical information of inventors, assignees and regions | C1: Inventor co-occurrence frequency | The deviation of the number of shared inventors of a pair of technology classes from the expected value under the hypothesis that diversification patterns are random. |
| | | C2: Organization co-occurrence frequency | The deviation of the number of shared inventing organizations of a pair of technology classes from the expected value under the hypothesis that diversification patterns are random. |
| | | C3: Country co-occurrence frequency | The deviation of the number of shared inventing countries of a pair of technology classes from the expected value under the hypothesis that diversification patterns are random. |
| D: Frequency for technology fields to share same patents | Information of multiple classes assigned to the same patent | D1: Normalized co-classification | The number of shared patents of a pair of technology classes, normalized by the number of all unique patents in both classes. |
| | | D2: Co-classification cosine similarity | The cosine of the angle of the two vectors representing two technology classes' distributions of shared patents with all other technology classes. |
| | | D3: Patent co-occurrence frequency | The deviation of the number of shared patents of a pair of technology classes from the expected value under the hypothesis that classes are randomly assigned to patents. |

The first group of measures (A1, A2 and A3) uses the information of backward citations (i.e. references) of patents, which represent the knowledge inputs to innovation activities, to measure either relatedness or similarity of knowledge bases or inputs of different classes.

A1. "Normalized co-reference": the count of shared references, normalized by the total count of all unique references of patents in a pair of classes, formulated as

$$\text{Co-Reference} = \frac{|C_i \cap C_j|}{|C_i \cup C_j|} \tag{1}$$

where $C_i$ and $C_j$ are the numbers of backward citations (i.e., references) of patents in technology classes $i$ and $j$; $|C_i \cap C_j|$ is the number of patents referenced in both technology classes $i$ and $j$, and $|C_i \cup C_j|$ is the total number of unique patents referenced in both technology classes $i$ and $j$, respectively. It is also known as the Jaccard index (Jaccard, 1901).

A2. "Class-to-class cosine similarity": the cosine of the angle of the two vectors representing two technology classes' distributions of citations into all patent classes (Leydesdorff, 2007), formulated as

$$\text{Cosine}(i,j) = \frac{\sum_k C_{ik} C_{jk}}{\sqrt{\sum_k C_{ik}^2} \sqrt{\sum_k C_{jk}^2}} \tag{2}$$

where $C_{ij}$ denotes the number of citations referred from patents in technology class $i$ to the patents in technology class $j$; $k$ belongs to all the technology classes. The cosine value is between [0,1] and indicates the similarity of the knowledge bases of two fields.

A3. "Class-to-patent cosine similarity": the cosine of the angle of the two vectors representing two technology classes' distributions of citations into specific unique patents instead of aggregated classes. The same formula as (2) applies, but $C_{ij}$ now denotes the number of citations of all patents in class $i$ to the specific patent $j$. Measure A3 has a better resolution than measure A2, whereas computation is slightly more complex.

The next two groups of measures, B1-B3 and C1-C3, similarly utilize the patent information related to successful inventive behaviors of different types of agents (inventors, organizations and countries) in terms of which classes their patents are assigned in. These measures generally indicate the likelihood or frequency that innovation agents diversify across a pair of technology fields. We separate them into two groups, B and C, due to the difference in their mathematical formulas.

B1. "Inventor diversification likelihood": minimum of the pairwise conditional probabilities ($R_{ij}$) of an inventor having strong inventing records in one class, given that this person also has strong inventing records in the other

$$R_{ij} = \min\{Prob(\text{RTA}_{c,i} | \text{RTA}_{c,j}), Prob(\text{RTA}_{c,j} | \text{RTA}_{c,i})\} \tag{3}$$

where $\text{RTA}_{c,i}$ and $\text{RTA}_{c,j}$ denotes inventor $c$'s revealed technological advantage in technology class $i$ and $j$.

$$\text{RTA}_{c,i} = \frac{x(c,i)}{\sum_i x(c,i)} \bigg/ \frac{\sum_c x(c,i)}{\sum_{c,i} x(c,i)} \tag{4}$$

where $x(c,i)$ is the number of patents held by an inventor $c$ in technology class $i$; $\sum_i x(c,i)$ is the number of patents held by an inventor $c$ in all technology classes; $\sum_c x(c,i)$ is the total number of patents held by all inventors $c$ in class $i$; $\sum_{c,i} x(c,i)$ is the total number of patents in the observed data. $\text{RTA}_{c,i}$ is an indicator of the relative inventive capacity of inventor $c$ in class $i$. $\text{RTA}_{c,i} > 1$ means inventor $c$ has more patents in class $i$ as a share of the inventor's total patents than an "average" inventor; otherwise, if $\text{RTA}_{c,i} \leq 1$. The mathematical formulation is adopted from Hidalgo et al. (2007). A high $R_{ij}$ value indicates a higher likelihood for an inventor to leverage knowledge across technology fields $i$ and $j$ for innovation, or to diversify personal inventive activities across fields $i$ and $j$. To calculate this measure, we use the unique inventor identifiers from the Institute for Quantitative Social Science at Harvard University (Li et al., 2014).

B2. "Organization diversification likelihood": the formula is the same as the "inventor diversification likelihood" above (Eq. 3 and Eq. 4), except that the agent is now an "organization", which is often a company, university, or public R&D institute. The organizations are identified using the "unique assignee" identifiers created by the National Bureau of Economics Research (NBER) (Hall et al., 2001).

B3. "Country diversification likelihood": the formula is the same as the "inventor diversification likelihood" above (Eq. 3 and Eq. 4), except that the agent is now a country.

Measures C1, C2 and C3 employ the form of a measure, which Teece et al. (1994) first proposed to indicate the relatedness between industries, following the "survivor principle" in economics. Applied to the context of technology classes, the measure compares the empirically observed frequency of co-occurrences of a pair of technology classes in the patenting records of the same inventors, organizations or countries, to the expected frequency in a random co-occurrence situation controlled for the sizes of technology classes.

C1: "Inventor co-occurrence frequency": the deviation of the empirically observed number of inventors occurring in a pair of technology classes from the value that would be expected when technology classes are randomly assigned to inventors. The formula is,

$$r_{ij} = \frac{O_{ij} - \mu_{ij}}{\sigma_{ij}} \tag{5}$$

where $O_{ij}$ is the observed number of inventors active in both technology classes $i$ and $j$, i.e. the count of inventor-level occurrences of patent technology classes $i$ and $j$; $\mu_{ij}$ and $\sigma_{ij}$ are the mean and variance of the expected number of inventors active in both classes *i* and *j*, given by a hyergeometric distribution. The hypoergeometric distribution defines $\mu_{ij}$ and $\sigma_{ij}$ as

$$\mu_{ij} = \frac{N_i N_j}{T} \tag{6}$$

$$\sigma^2 = \mu_{ij} (\frac{T-N_i}{T})(\frac{T-N_j}{T-1}) \tag{7}$$

where *T* is the total number of inventors having two or more technology classes; $N_i$ and $N_j$ are the number of inventors empirically observed in technology *i* and *j*, respectively. Thus $r_{ij}$ is analogous to a *t*-statistic. It indicates that, when the actual number $O_{ij}$ observed between classes *i* and *j* greatly exceeds the expected number $\mu_{ij}$, these two technology fields are considered highly proximate.

This measure controls for the effect of the sizes of technology classes on inventor appearances in them. If technology classes have many inventors, i.e. $N_i$ (or $N_j$) is large, the chance for inventors in $N_i$ (or $N_j$) to be active in class *j* would be high, even if classes *i* and *j* are distant. When $N_i$ or $N_j$ is small, one would not expect to see many co-occurrences even if classes *i* and *j* are highly proximate. $\mu_{ij}$ estimates the co-occurrence frequency resulting only from the size effect but not from the effect of knowledge distance. Therefore, $r_{ij}$ extracts the information in $O_{ij}$ about actual distance.

C2. "Organization co-occurrence frequency": the formula is the same as "inventor co-occurrence frequency" above (Eq. 5-7), except that the agent is now an "organization", which is often a company, university, or public R&D agency.

C3. "Country co-occurrence frequency": the formula is the same as "inventor diversification likelihood" above (Eq. 5-7), except that the agent is now a country.

The last group of measures (D1, D2 and D3) uses the information of the co-classifications of patents to quantify the co-occurrences of a pair of technology classes in the same patents. Co-classification means that a patent is assigned to more than one class. Patent examiners based on their assessments of the inventions carry out the assignment activity.

D1. "Normalized co-classification": the count of shared patents, normalized by the total count of unique patents in a pair of classes, formulated as

$$\text{Co-Classification} = \frac{|N_i \cap N_j|}{|N_i \cup N_j|} \tag{8}$$

where $N_i$ and $N_j$ are the number of patents in technology classes *i* and *j*, respectively; $|N_i \cap N_j|$ is the number of shared patents in both technology classes *i* and *j*, and $|N_i \cup N_j|$ is the total number of unique patents in both technology classes *i* and *j*.

D2. "Co-classification cosine similarity": the cosine of the angle of the two vectors representing two technology classes' distributions of patents shared with all other technology classes, formulated as

$$\text{Cosine}(i, j) = \frac{\sum_k O_{ik} O_{jk}}{\sqrt{\sum_k O_{ik}^2} \sqrt{\sum_k O_{jk}^2}} \qquad (9)$$

where $O_{ij}$ is the number of shared patents in both technology classes $i$ and $j$.

D3: "Patent co-occurrence frequency": the deviation of the empirically observed number of patents occurring in a pair of technology classes from the value that would be expected when technology classes are randomly assigned to patents. Its formulas are the same as those for inventor, organization and country co-occurrence frequencies, i.e. Eq. (5)-(7). But here the variables are given new meanings: $O_{ij}$ is the number of shared patents in both classes $i$ and $j$; $T$ is the total number of patents having two or more technology classes; $N_i$ and $N_j$ are the number of patents in classes $i$ and $j$, respectively. D3 concerns the frequency of patents being assigned to a pair of classes, differing from C1, C2 and C3 that concern the frequency of innovation agents being active in a pair of classes.

By far we have introduced 12 measures to be analyzed in Section 4. While some measures (such as A2, B3, D2 and D3) have appeared in prior studies of technology networks, the others (such as A1, A3, B1, B2, C1, C2, C3 and D1) are new. A1 and D1 use the formula of well-established *Jaccard index*, but are new to the literature on patent network mapping. While A3 employs the cosine formula of A2 that has been used to construct patent technology networks (Leydesdorff et al., 2014; Kay et al., 2014), it is new in that it considers the distribution of citations to unique patents, rather than classes, in order to improve measurement resolution. The formula for B1, B2 and B3 first appeared in the studies on the diversification of countries or regions in the product space based on export product data (Hidalgo et al., 2007; Boschma et al., 2014). To our best knowledge, the present paper is the first to apply this formula to the analysis levels of inventors and inventive organizations in the context of technology and patent classes. Thus we consider B1 (*inventor diversification likelihood*) and B2 (*organization diversification likelihood*) are new to the literature. Likewise, the formula of C1, C2 and C3 was first developed to measure industry relatedness (Teece et al., 1994). Here, it is the first time that the formula is used to measure the frequency of innovation agents having patents in pairs of patent technology classes.

*3.3 Strategy of comparison*

Because these distance measures deal with the same underline structure of the technology space, the overall network map structures resulting from them are expected to be similar. The maps that exhibit the highest structural similarities with all other maps are the best representatives out of this set of alternative maps to represent the underlining technology space. Therefore, after the networks of the 121 IPC classes are constructed by using the twelve distance measures, we investigate their pairwise correlations/similarities in terms of network structural properties, for example, weights of corresponding edges and centralities of corresponding vertices in different networks.

To calculate the centrality of each vertex, we employ the two most commonly used network centrality metrics in graph theory, because of their applicability to weighted undirected networks. One is *degree centrality*, which is the sum of the weights of the edges connected to the focal vertex. The other is *eigenvector centrality*, which is the value of the focal vertex's respective element in the dominant eigenvector of the adjacency matrix of the network (Newman, 2005). We also assess the correlation between vertex centralities and the indicators of "importance" of technology classes (e.g., total numbers of patents and forward citation counts of patents in a class), in the 12 technology networks.

## 4. Results

Before comparing different types of networks, we first examine the over-time changes of each of them. Table 2 reports the Pearson coefficients of correlations of the weights of corresponding edges in different time periods for each network. It is shown that the correlations between networks of the same type but in different time periods are generally high (in most cases, higher than 0.9), indicating that network changes over time are fairly slow regardless of the choices of distance measures. Such observations are consistent with the prior study of Hinze et al. (1997) that also suggested the stability of the network of technology fields.

In the meantime, these two networks based on measures of the likelihood and frequency of country-level cross-field diversification (B3 and C3) are the least stable over time, as indicated by their lowest correlation coefficients in the range of 0.5~0.6. For all measures, the network based on data for the longest time period (1976 to 2006) is the most correlated with all other networks constructed using data for single decades (e.g. 1977 to 1986, 1987 to 1996, 1997 to 2006). Therefore, in later analyses we focus on the networks constructed using our total patent data from 1976 to 2006, to have the most representative empirical approximation of the distance between technology fields.

**Table 2** Correlation coefficients of edge weights in the same technology network for different time periods

| A1. Normalized co-reference | 1977-1986 | 1987-1996 | 1997-2006 | 1976-2006 |
|---|---|---|---|---|
| 1977-1986 | 1.000 | | | |
| 1987-1996 | 0.952 | 1.000 | | |
| 1997-2006 | 0.894 | 0.951 | 1.000 | |
| 1976-2006 | 0.937 | 0.977 | 0.987 | 1.000 |
| A2. Class-class cosine | 1977-1986 | 1987-1996 | 1997-2006 | 1976-2006 |
| 1977-1986 | 1.000 | | | |
| 1987-1996 | 0.896 | 1.000 | | |
| 1997-2006 | 0.817 | 0.882 | 1.000 | |
| 1976-2006 | 0.874 | 0.921 | 0.987 | 1.000 |
| A3. Class-patent cosine | 1977-1986 | 1987-1996 | 1997-2006 | 1976-2006 |
| 1977-1986 | 1.000 | | | |
| 1987-1996 | 0.877 | 1.000 | | |
| 1997-2006 | 0.712 | 0.812 | 1.000 | |
| 1976-2006 | 0.817 | 0.913 | 0.973 | 1.000 |
| B1. Inventor diversification likelihood | 1977-1986 | 1987-1996 | 1997-2006 | 1976-2006 |
| 1977-1986 | 1.000 | | | |
| 1987-1996 | 0.943 | 1.000 | | |
| 1997-2006 | 0.846 | 0.916 | 1.000 | |
| 1976-2006 | 0.928 | 0.973 | 0.972 | 1.000 |
| B2. Organization diversification likelihood | 1977-1986 | 1987-1996 | 1997-2006 | 1976-2006 |
| 1977-1986 | 1.000 | | | |
| 1987-1996 | 0.929 | 1.000 | | |
| 1997-2006 | 0.864 | 0.938 | 1.000 | |
| 1976-2006 | 0.935 | 0.969 | 0.969 | 1.000 |
| B3. Country diversification likelihood | 1977-1986 | 1987-1996 | 1997-2006 | 1976-2006 |
| 1977-1986 | 1.000 | | | |
| 1987-1996 | 0.592 | 1.000 | | |
| 1997-2006 | 0.547 | 0.633 | 1.000 | |
| 1976-2006 | 0.654 | 0.752 | 0.819 | 1.000 |
| C1. Inventor co-occurrence frequency | 1977-1986 | 1987-1996 | 1997-2006 | 1976-2006 |
| 1977-1986 | 1.000 | | | |
| 1987-1996 | 0.950 | 1.000 | | |
| 1997-2006 | 0.879 | 0.942 | 1.000 | |
| 1976-2006 | 0.939 | 0.978 | 0.979 | 1.000 |
| C2. Organization co-occurrence frequency | 1977-1986 | 1987-1996 | 1997-2006 | 1976-2006 |
| 1977-1986 | 1.000 | | | |
| 1987-1996 | 0.938 | 1.000 | | |
| 1997-2006 | 0.886 | 0.932 | 1.000 | |
| 1976-2006 | 0.938 | 0.965 | 0.968 | 1.000 |
| C3. Country co-occurrence frequency | 1977-1986 | 1987-1996 | 1997-2006 | 1976-2006 |
| 1977-1986 | 1.000 | | | |
| 1987-1996 | 0.541 | 1.000 | | |
| 1997-2006 | 0.483 | 0.687 | 1.000 | |
| 1976-2006 | 0.606 | 0.746 | 0.889 | 1.000 |
| D1. Normalized co-classification | 1977-1986 | 1987-1996 | 1997-2006 | 1976-2006 |
| 1977-1986 | 1.000 | | | |
| 1987-1996 | 0.922 | 1.000 | | |
| 1997-2006 | 0.654 | 0.741 | 1.000 | |
| 1976-2006 | 0.896 | 0.947 | 0.904 | 1.000 |
| D2. Co-classification cosine | 1977-1986 | 1987-1996 | 1997-2006 | 1976-2006 |
| 1977-1986 | 1.000 | | | |
| 1987-1996 | 0.873 | 1.000 | | |
| 1997-2006 | 0.784 | 0.874 | 1.000 | |
| 1976-2006 | 0.912 | 0.954 | 0.948 | 1.000 |
| D3. Patent co-occurrence frequency | 1977-1986 | 1987-1996 | 1997-2006 | 1976-2006 |
| 1977-1986 | 1.000 | | | |
| 1987-1996 | 0.923 | 1.000 | | |
| 1997-2006 | 0.810 | 0.890 | 1.000 | |
| 1976-2006 | 0.931 | 0.973 | 0.957 | 1.000 |

To develop a general intuition about technology network structures, we visualize these twelve networks, using VOSviewer that was initially created to visualize bibliometric networks. Leydesdorff et al. (2014) showed that it also provides good visualizations of patent technology networks. As an example, Fig. 2 visualizes the technology network constructed by using the measure of *inventor diversification likelihood* (B1). This network is also the background map used in Fig. 1 to locate the specific knowledge positions of Google and to reveal its innovation directions or diversification paths by overlaying.

These networks are almost fully connected, but most of the edges have extremely small values, indicating long distance between most fields. Therefore, to visually reveal its main structure, we filter the network to contain only the *maximum spanning tree*[4] as the skeleton plus the strongest edges, which together make the total number of edges be twice that of the vertices. Hildago et al. (2007) suggested this threshold of edge filtering as a rule of thumb for good network visualization. Fig. 2 is such a filtered network. It exhibits a heterogeneous structure, with six communities of technology fields identified by the Louvain community detection method (Blondel et al., 2008). The heterogeneity, instead of homogeneity, of the structure of technology networks justifies it as a good lens or protocol for the comparison of alterative networks.

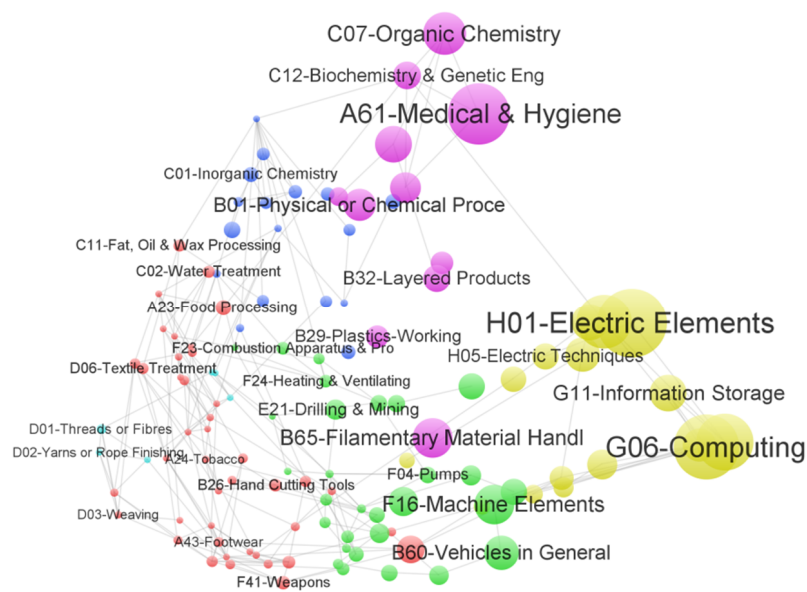

**Fig. 2** The technology network using the inventor diversification likelihood as distance measure. Vertex sizes correspond to the total patent counts in respective IPC patent classes; vertex colors denote different communities.

---

[4] A maximum spanning tree (MST) only keeps those strongest edges that minimally connect the network into a tree. The MST for the network in Fig. 2 contains the strongest 120 edges that connect all 121 vertices into a tree.

*4.1 Correlation of edge weights between different networks*

To compare the structures of the 12 networks, we first investigate the Pearson correlation coefficient between the edge weights of corresponding pairs of technology classes in different networks (see Table 3). In Table 3, we underline the highest correlation coefficient between each measure and any other measures, and also report the average of the correlation coefficients of one measure with all other 11 measures. In respective categories (A, B, C and D), the networks using the measures of *normalized co-reference* (A1), *inventor diversification likelihood* (B1), *inventor co-occurrence frequency* (C1) and *co-classification* (D1) are the most correlated with all other types of networks. This result suggests, A1, B1, C1 and D1 lead to the most representative networks in their respective groups.

**Table 3** Correlation coefficients of edge weights between technology networks (1976-2006)

|   | A1 | A2 | A3 | B1 | B2 | B3 | C1 | C2 | C3 | D1 | D2 | D3 |
|---|---|---|---|---|---|---|---|---|---|---|---|---|
| A1 |  | 0.541 | 0.857 | 0.915 | 0.785 | 0.178 | 0.792 | 0.723 | 0.145 | 0.736 | 0.453 | 0.587 |
| A2 | 0.541 |  | 0.661 | 0.482 | 0.414 | 0.021 | 0.592 | 0.460 | -0.119 | 0.454 | 0.446 | 0.490 |
| A3 | 0.857 | 0.661 |  | 0.814 | 0.664 | 0.142 | 0.775 | 0.622 | 0.083 | 0.734 | 0.350 | 0.635 |
| B1 | 0.915 | 0.482 | 0.814 |  | 0.840 | 0.205 | 0.827 | 0.737 | 0.191 | 0.836 | 0.398 | 0.666 |
| B2 | 0.785 | 0.414 | 0.664 | 0.840 |  | 0.389 | 0.703 | 0.866 | 0.371 | 0.609 | 0.431 | 0.522 |
| B3 | 0.178 | 0.021 | 0.142 | 0.205 | 0.389 |  | 0.259 | 0.288 | 0.428 | 0.139 | 0.171 | 0.228 |
| C1 | 0.792 | 0.592 | 0.775 | 0.827 | 0.703 | 0.259 |  | 0.714 | 0.134 | 0.736 | 0.451 | 0.898 |
| C2 | 0.723 | 0.460 | 0.622 | 0.737 | 0.866 | 0.288 | 0.714 |  | 0.413 | 0.529 | 0.489 | 0.522 |
| C3 | 0.145 | -0.119 | 0.083 | 0.191 | 0.371 | 0.428 | 0.134 | 0.413 |  | 0.085 | 0.133 | 0.112 |
| D1 | 0.736 | 0.454 | 0.734 | 0.836 | 0.609 | 0.139 | 0.736 | 0.529 | 0.085 |  | 0.238 | 0.735 |
| D2 | 0.453 | 0.446 | 0.350 | 0.398 | 0.431 | 0.171 | 0.451 | 0.489 | 0.133 | 0.238 |  | 0.279 |
| D3 | 0.587 | 0.490 | 0.635 | 0.666 | 0.522 | 0.228 | 0.898 | 0.522 | 0.112 | 0.735 | 0.279 |  |
| Average | 0.610 | 0.404 | 0.576 | 0.628 | 0.599 | 0.223 | 0.626 | 0.578 | 0.180 | 0.530 | 0.349 | 0.516 |

\* Distance measures: (A1) normalized co-reference; (A2) class-to-class cosine similarity; (A3) class-to-patent cosine similarity; (B1) inventor diversification likelihood; (B2) organization diversification likelihood; (B3) country diversification likelihood; (C1) inventor co-occurrence frequency; (C2) organization co-occurrence frequency; (C3) country co-occurrence frequency; (D1) normalized co-classification; (D2) co-classification cosine similarity; (D3) patent co-occurrence frequency.

In particular, among all the pairwise correlations, the correlation coefficient (=0.915) for the pair of A1 and B1 is the highest. This may suggest a strong effect of the technological distance of a pair of fields on the likelihood for inventors to diversify across fields or to combine knowledge of these fields to generate new inventions. In the meantime, this effect of knowledge distance on diversification patterns is lesser for organizations and the least for countries. The networks based on *country diversification likelihood* (B3) and *country co-occurrence frequency* (C3) are the least correlated with networks in group A based on knowledge distance measures as well as all other networks. That is, the *survivor principle* may not explain technology diversification patterns of *countries*. A country may survive many inefficient diversification choices and behaviors across distant fields, as long as their resources are abundant enough to ensure survival. In contrast, individual inventors must learn and master relevant knowledge of a technology field in order to invent there. Thus the diversification patterns of inventors, as measured by B1 and C1, are strongly constrained by the knowledge distance of different fields, as measured by A1.

We further plot and visually compare the distributions of edges by weights of the 12 network maps in Fig. 3. The networks using group A measures and the networks using B1 and B2 as well as D1 measures exhibit negative exponential distributions. In particular, the high skewness of the A1, A2 and A3 networks indicates that very most technology fields are indeed highly distant from one another. Thus, the likelihood for inventors to diversify across most pairs of technology fields must be also limited. This is reflected in the similarly skewed distribution of edges in the network using *inventor diversification likelihood* (B1).

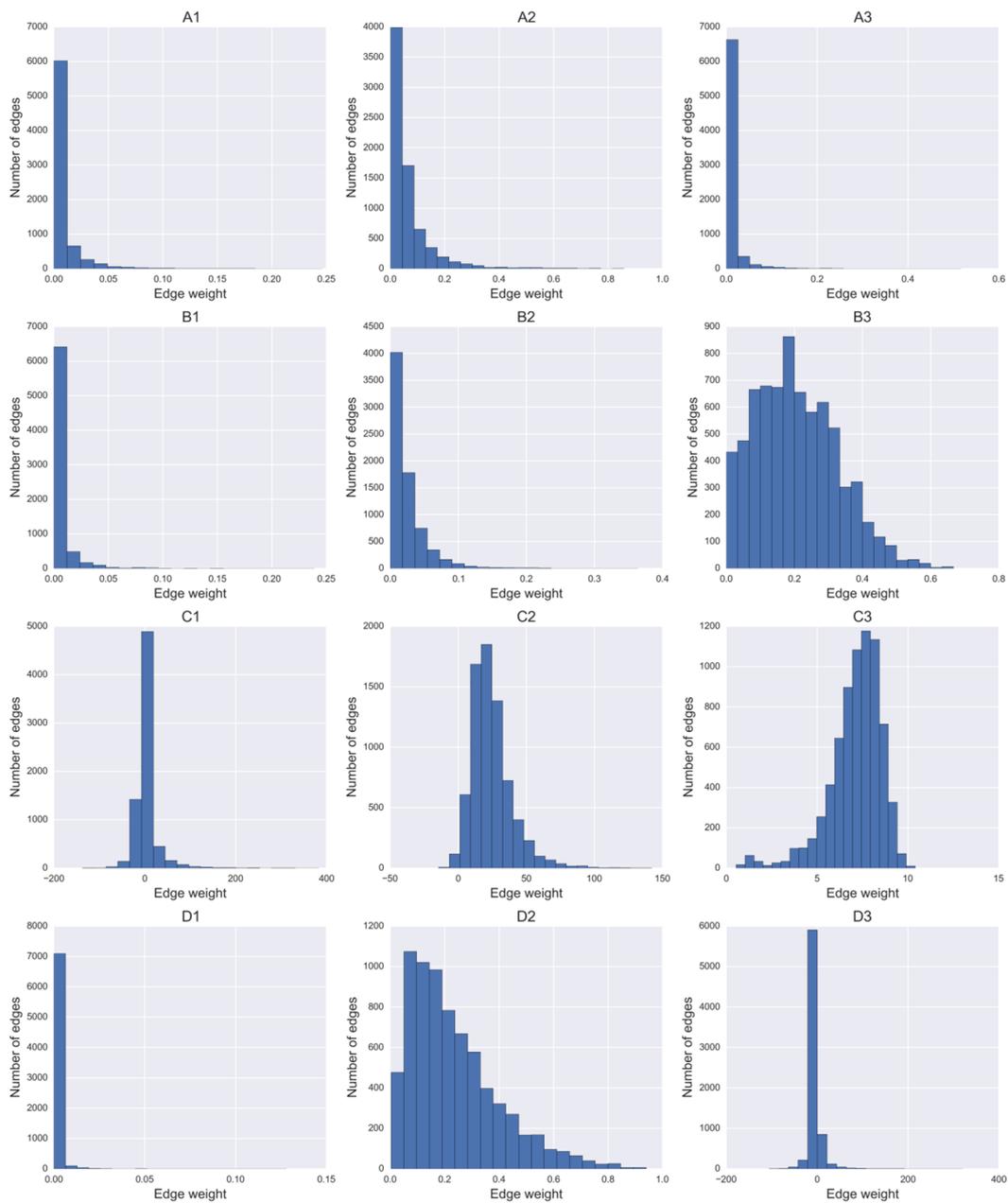

**Fig. 3** Distributions of edges by weights of the technology networks using 12 alternative measures.

In group B, as we expand the scope of the innovation agents from inventors (B1) to organizations (B2) and then to countries (B3), the mean values of inter-field distance decreases (i.e. proximity increases) and the skewness of the distribution decreases. This increasing normality from B1 to B3 may also suggest that *countries* may make more *normal* decisions of cross-field diversifications, without being strongly constrained by inter-field knowledge distance whose distributions are highly skewed. In addition, the distributions of C1, C2, C3 and D3 all exhibit the form of normal distributions, despite varied skewness. This similarity may result from their shared mathematical formation that normalizes empirical observations with corresponding random scenarios.

4.2 *Correlation of vertex centralities between different networks*

We also investigate the correlation between degree and eigenvector centralities of the same set of 121 vertices in different networks. In Tables 4(a) and 4(b), we underline the highest correlation coefficients between each measure and any other measures, and also report the average of the correlation coefficients of each measure with all other 11 measures. For both types of centrality metrics, within group A, the networks using *normalized co-reference* (A1) and *class-to-patent cosine similarity* (A3) are far more correlated with all other networks than *class-to-class cosine similarity* (A2), although A2 is popularly used in the literature (Leydesdorff et al., 2014; Kay et al., 2014).

**Table 4(a)** Correlations of vertex degree centrality between technology networks (1976-2006)

|  | A1 | A2 | A3 | B1 | B2 | B3 | C1 | C2 | C3 | D1 | D2 | D3 |
|---|---|---|---|---|---|---|---|---|---|---|---|---|
| A1 |  | 0.236 | <u>0.914</u> | <u>0.910</u> | 0.539 | -0.179 | 0.101 | 0.496 | 0.047 | 0.837 | 0.338 | -0.521 |
| A2 | 0.236 |  | 0.474 | 0.168 | 0.051 | -0.369 | 0.132 | 0.119 | -0.334 | 0.277 | 0.131 | -0.107 |
| A3 | 0.914 | <u>0.474</u> |  | 0.833 | 0.502 | -0.192 | 0.201 | 0.453 | -0.003 | <u>0.844</u> | 0.265 | -0.419 |
| B1 | <u>0.910</u> | 0.168 | 0.833 |  | 0.760 | -0.062 | 0.295 | 0.709 | 0.241 | 0.819 | 0.273 | -0.361 |
| B2 | 0.539 | 0.051 | 0.502 | 0.760 |  | 0.380 | 0.493 | <u>0.876</u> | 0.514 | 0.493 | 0.156 | 0.018 |
| B3 | -0.179 | -0.369 | -0.192 | -0.062 | 0.380 |  | 0.242 | 0.149 | 0.435 | -0.126 | -0.064 | 0.309 |
| C1 | 0.101 | 0.132 | 0.201 | 0.295 | 0.493 | 0.242 |  | 0.496 | 0.387 | -0.049 | 0.285 | <u>0.734</u> |
| C2 | 0.496 | 0.119 | 0.453 | 0.709 | <u>0.876</u> | 0.149 | 0.496 |  | <u>0.560</u> | 0.390 | 0.230 | 0.033 |
| C3 | 0.047 | -0.334 | -0.003 | 0.241 | 0.514 | <u>0.435</u> | 0.387 | 0.560 |  | -0.029 | 0.194 | 0.286 |
| D1 | 0.837 | 0.277 | 0.844 | 0.819 | 0.493 | -0.126 | -0.049 | 0.390 | -0.029 |  | 0.043 | -0.632 |
| D2 | 0.338 | 0.131 | 0.265 | 0.273 | 0.156 | -0.064 | 0.285 | 0.230 | 0.194 | 0.043 |  | 0.044 |
| D3 | -0.521 | -0.107 | -0.419 | -0.361 | 0.018 | 0.309 | <u>0.734</u> | 0.033 | 0.286 | -0.632 | 0.044 |  |
| Average | 0.338 | 0.071 | 0.352 | 0.417 | 0.435 | 0.048 | 0.302 | 0.410 | 0.209 | 0.261 | 0.172 | -0.056 |

**Table 4(b)** Correlations of vertex eigenvector centrality between technology networks (1976-2006)

|  | A1 | A2 | A3 | B1 | B2 | B3 | C1 | C2 | C3 | D1 | D2 | D3 |
|---|---|---|---|---|---|---|---|---|---|---|---|---|
| A1 |  | 0.285 | <u>0.826</u> | <u>0.891</u> | 0.418 | -0.278 | -0.041 | 0.472 | -0.001 | <u>0.391</u> | 0.293 | 0.046 |
| A2 | 0.285 |  | 0.502 | 0.284 | 0.053 | -0.41 | -0.167 | 0.159 | -0.338 | 0.165 | 0.101 | -0.061 |
| A3 | 0.826 | <u>0.502</u> |  | 0.801 | 0.437 | -0.17 | -0.342 | 0.44 | 0.038 | 0.368 | 0.199 | -0.066 |
| B1 | <u>0.891</u> | 0.284 | 0.801 |  | 0.671 | -0.171 | -0.138 | 0.701 | 0.164 | 0.267 | 0.234 | 0.106 |
| B2 | 0.418 | 0.053 | 0.437 | 0.671 |  | 0.39 | -0.212 | <u>0.872</u> | 0.478 | 0.014 | 0.09 | 0.112 |
| B3 | -0.278 | -0.41 | -0.17 | -0.171 | 0.39 |  | -0.307 | 0.128 | 0.411 | -0.088 | -0.088 | -0.123 |
| C1 | -0.041 | -0.167 | -0.342 | -0.138 | -0.212 | -0.307 |  | 0.004 | 0.101 | -0.619 | 0.281 | <u>0.786</u> |
| C2 | 0.472 | 0.159 | 0.44 | 0.701 | <u>0.872</u> | 0.128 | 0.004 |  | <u>0.534</u> | -0.088 | 0.198 | 0.282 |
| C3 | -0.001 | -0.338 | 0.038 | 0.164 | 0.478 | <u>0.411</u> | 0.101 | 0.534 |  | -0.296 | 0.209 | 0.292 |
| D1 | 0.391 | 0.165 | 0.368 | 0.267 | 0.014 | -0.088 | -0.619 | -0.088 | -0.296 |  | -0.246 | -0.868 |
| D2 | 0.293 | 0.101 | 0.199 | 0.234 | 0.09 | -0.088 | 0.281 | 0.198 | 0.209 | -0.246 |  | 0.378 |

| D3 | 0.046 | -0.061 | -0.066 | 0.106 | 0.112 | -0.123 | <u>0.786</u> | 0.282 | 0.292 | -0.868 | <u>0.378</u> | |
| Average | 0.300 | 0.052 | 0.276 | 0.346 | 0.302 | -0.064 | -0.059 | 0.337 | 0.145 | -0.091 | 0.150 | 0.080 |

Within group B, the networks using *inventor diversification likelihood* (B1) and *organization diversification likelihood* (B2) are far more correlated with other networks, than *country diversification likelihood* (B1). In group C, the network using *organization co-occurrence frequency* (C2) is the most correlated with other types of networks, in contrast to the analysis of edge weight correlations (see Table 3) that suggests the network using *inventor co-occurrence frequency* (C1) is the most correlated one from group C. The results of group D networks are mixed across the analyses of degree and eigenvector centralities, and the correlations of group D networks with other networks are generally weak.

In the meantime, we find a few negative correlation coefficients in Table 4. The negative correlations between group A networks (based on knowledge distance) and *country diversification likelihood* (B3) and *country co-occurrence frequency* (C3) may suggest that, it is more often for countries to diversify into or combine knowledge of less central fields in the technology space, implying broad exploration.

In brief, based on the analysis of pairwise network vertex centrality correlations, A1 and A3, B1, B2, and C2 lead to the most representative network maps.

*4.3 Correlations of vertex centrality, popularity and impact in different networks*

The relative network positions of technology fields in the total technology space may affect their relative popularity (i.e. attracting innovation activities) and impact (i.e. influencing future innovation activities). For instance, for the technology classes that are strongly connected to many other classes (i.e. high centrality in the technology network), innovation agents from many other fields may enjoy the ease to diversify into them, resulting in a large number of patents. Innovation agents in highly central fields can also see many innovation opportunities via leveraging and recombining knowledge from many other fields strongly connected to their own. Thus, the relative network positions of technology classes in terms of centralities may predict their relative importance. Herein, we compare such predictabilities of the 12 networks.

We focus on two indicators of the importance of a technology field: popularity and impact. Popularity of a technology field is measured by the total number of all patents in its patent class from 1976 to 2006. Impact of a field is measured by the number of total forward citations of all patents in the field, i.e. the number of later patents that cite a focal patent.

Forward citation count of a patent has been found to be an effective indicator of the economic impact of the patent in a number of prior studies (Albert et al., 1991; Hall et al., 2005; Harhoff et al., 1999; Lee et al., 2007; Trajtenberg, 1990). We calculate Spearman rank correlations between vertex centralities and popularity/impact indicators of the 121 technology classes in each of the 12 networks, because it is more appropriate for the purpose of investigating if more central vertices are also more important ones. Results are shown in Table 5.

**Table 5**. Rank correlations of vertex centralities, forward citation counts and patent counts in different networks.

| Networks | Degree Centrality | | Eigenvector Centrality | |
|---|---|---|---|---|
| | # of future citations | # of patents | # of future citations | # of patents |
| A1 | 0.859 | 0.857 | 0.875 | 0.874 |
| A2 | -0.036 | -0.036 | -0.027 | -0.025 |
| A3 | 0.654 | 0.657 | 0.546 | 0.550 |
| B1 | 0.776 | 0.773 | 0.782 | 0.782 |
| B2 | 0.300 | 0.296 | 0.281 | 0.278 |
| B3 | -0.279 | -0.278 | -0.277 | -0.275 |
| C1 | -0.024 | -0.032 | 0.290 | 0.275 |
| C2 | 0.284 | 0.280 | 0.292 | 0.289 |
| C3 | -0.126 | -0.123 | -0.129 | -0.125 |
| D1 | 0.738 | 0.741 | 0.694 | 0.701 |
| D2 | 0.312 | 0.304 | 0.310 | 0.300 |
| D3 | -0.841 | -0.845 | 0.547 | 0.531 |

Among all 12 networks, the one using *normalized co-reference* (A1) yields the highest correlations of vertices between their network centralities and total patent counts as well as forward citation counts. A1 is followed by the networks using *inventor diversification likelihood* (B1), *normalized co-classification* (D1) and *class-to-patent cosine similarity* (A3), in terms of the predictability of vertex centralities on actual importance. In these networks, it is highly likely that more central technology classes are also more popular and impactful ones. None of other networks than A1, B1, D1 and A3 provides a correlation coefficient greater than 0.5.

*4.4 Brief summary of findings*

We have analyzed different kinds of correlations in terms of edge weights, vertex centrality and importance, among the 12 technology networks, in order to explore how similar each network is with other networks. We paid special attention to the networks that have the highest overall correlations with all other networks. Specifically, the measures of *normalized co-reference* (A1) and *inventor diversification likelihood* (B1) lead to network maps that are consistently the most correlated with other networks, in terms of edge and vertex properties of the networks, and also provide the best correlations between network centrality and importance indicators of technology classes. These networks are the best representatives of the technology space from this set of 12 candidate alternatives compared in this study.

A few findings on additional measures are noteworthy. First, in group A, *class-to-class cosine similarity* (A2) performs the worst in various correlation analyses, although it is popularly used for constructing network maps of patent technology classes in the literature. *Class-to-patent cosine similarity* (A3) follows the same formula of A2 but uses a higher resolution of data, and is highly correlated with A1. In some cases of our analysis, A3 is no worse than A1. Second, the networks using *country diversification likelihood* (B3) and *country co-occurrence frequency* (C3) are the least correlated with the networks using measures in group A. Furthermore, *inventor co-occurrence frequency* (C1), *organization co-occurrence frequency* (C2) and *normalized co-classification* (D1) perform well in some of our correlation analyses, but not always.

## 5. Conclusion

This paper contributes to the research to develop and analyze patent technology network maps to explore technology diversification and knowledge combination opportunities, and thus support technology forecasting and road mapping practices. A main challenge to developing reasonable technology network maps is the ambiguity in the choice of measures of the distance between various technology fields in the total technology space. To address this challenge, we have proposed a strategy to assess and compare alternative measures through analyzing the overall structures of their resulting networks, because the structures of technology networks condition the strategic insights that can be potentially drawn for road-mapping and forecasting analyses but are affected by the choices of distance measures.

We implemented the strategy in a comparative analysis of twelve distance measures, by correlating the edge and vertex properties of their resulting networks based on network structural properties. These twelve measures were chosen to represent the most common types of distance measures that have been proposed in the literature. These measures had not been previously compared via a consistent quantitative methodology. Particularly, our analyses in all cases consistently suggest the measures of *normalized co-reference* and *inventor diversification likelihood* lead to network maps that are the most similar to all other maps. That is, these two measures lead to relatively the most representative technology maps in our comparative set of twelve.

The contribution of the present paper lies primarily in promising and demonstrating the strategy to assess alternative distance measures by analyzing the correlations or similarity of the overall network structures resulting from those measures. For a consistent comparison of alternative distance measures, we utilized IPC 3-digit classes to represent technological fields and operationalize the vertices in the networks. Moving forward, the same network-based comparative strategy and analysis that we have demonstrated here can be also be carried out on alternative definitions of vertices (representing technology fields) and categorizations of patents, such as USPC (United States Patent Classification), CPC (Cooperative Patent Classification), and the hybrid patent categories proposed by Kay et al. (2014) and Benson and Magee (2013; 2015)[5], to see which distance measures lead to the most representative maps. We consider this a promising avenue for future research.

Also, we only compared twelve distance measures. Although they are representative, additional measures exist in the literature. Researchers may also propose new types of measures in future. Therefore, future work may compare a broader set of existing and new measures. In addition, the present analyses only made use of a few simple graph theory metrics to analyze the correlations of the networks in terms of their structures, for demonstrating our proposed distance measure comparison strategy based on network analysis. More sophisticated network analysis methods can be applied to this kind of analysis. For instance, one can investigate and compare the community structure of alternative patent technology networks, using community detection algorithms. Eventually, we aim to investigate how to effectively use the technology network maps to explore knowledge combination opportunities for innovation and technology diversification directions of individual inventors, technology companies, cities and countries.

**Acknowledgement**

We thank Loet Leydesdorff and Duncan Kushnir for sharing a unique patent dataset, and Kristin Wood, Aditya Mathur, Christopher Magee, Jeffrey Alstott and Giorgio Truilzi for many insightful and stimulating discussions. This research was supported by SUTD-MIT International Design Centre and Singapore Ministry of Education Tier 2 Research Grants.

---

[5] Kay et al. (2014) regrouped patent classes in order to have more evenly distributed patents in hybrid classes for better visualization. Benson and Magee (2013; 2015)'s hybrid categories of patents aim to improve their completeness and relevancy in representing actual technologies from technologist perspective.